# The amplification of risk in experimental diffusion chains

Mehdi Moussaïd[1,2*], Henry Brighton[2], and Wolfgang Gaissmaier[2,3,4]

[1] Center for Adaptive Rationality, Max Planck Institute for Human Development, Lentzeallee 94, 14195 Berlin, Germany

[2] Center for Adaptive Behavior and Cognition, Max Planck Institute for Human Development, Lentzeallee 94, 14195 Berlin, Germany

[3] Harding Center for Risk Literacy, Max Planck Institute for Human Development, Lentzeallee 94, 14195 Berlin, Germany

[4] Department of Psychology, University of Konstanz, 7857 Konstanz Germany

* Corresponding author: mehdi.moussaid@gmail.com

## Abstract

Understanding how people form and revise their perception of risk is central to designing efficient risk communication methods, eliciting risk awareness, and avoiding unnecessary anxiety among the public. Yet public responses to hazardous events such as climate change, contagious outbreaks, and terrorist threats are complex and difficult to anticipate phenomena. While many psychological factors influencing risk perception have been identified in the past, it remains unclear how perceptions of risk change when propagated from one person to another, and what impact the repeated social transmission of perceived risk has at the population scale. Here, we study the social dynamics of risk perception by analysing how messages detailing the benefits and harms of a controversial antibacterial agent undergo change when passed from one person to the next in 10-subject experimental diffusion chains. Our analyses show that when messages are propagated through the diffusion chains, they tend to become shorter, gradually inaccurate, and increasingly dissimilar between chains. In contrast, the perception of risk is propagated with higher fidelity due to participants manipulating messages to fit their preconceptions, thereby influencing the judgments of subsequent participants. Computer simulations implementing this simple influence mechanism show that small judgment biases tend to become more extreme, even when the injected message contradicts preconceived risk judgments. Our results provide quantitative insights into the social amplification of risk perception, and can help policy makers better anticipate and manage the public response to emerging threats.

## Significance Statement

Public risk perception of hazardous events such as contagious outbreaks, terrorist attacks, and climate change are difficult to anticipate social phenomena. It is unclear how risk information will spread through social networks, how laypeople influence each other, and what social dynamics generate public opinion. We examine how messages detailing risks are transmitted from one person to another in experimental diffusion chains, and how people influence each other as they propagate this information. While the content of message is gradually lost over repeated social transmissions, subjective perceptions of risk propagate and amplify due to social influence. These results provide quantitative insights into the public response to risk, and understanding the formation of often unnecessary fears and anxieties.

# Introduction

Public perception of risk is often polarized, difficult to anticipate, and at odds with scientific evidence (1, 2). The risks associated with nuclear energy, genetically modified food, and nanotechnologies continue to elicit strong public reaction in contrast to the assessments of many experts, while policy makers often fail to influence the public perception of the risks associated with poor nutrition, a sedentary lifestyle, and over exposure to the sun (2). The ability to communicate risks to the public, but also the ability to anticipate the public response to these risks, has a substantial impact on society and is key to alleviating unnecessary public anxiety (3). While research into the psychological factors that influence the formation of risk perceptions (4–7) and the communication of risks (8, 9) is relatively well-developed, the study of risk perception is largely focused on the *individual*. Yet, risk judgments are formed in a *social* context: People frequently discuss the everyday risks they face with their friends, relatives, co-workers, and with unknown people over the Web. They observe and imitate the risky and self-protective behavior of their peers (10), and exchange opinions, sources of information, and behavioral recommendations through social media and online communication platforms. For example, during the H1N1 influenza outbreak in 2009, nearly half a million messages mentioning H1N1 vaccination were exchanged on Twitter, 20% of which explicitly expressed a positive or negative attitude toward vaccination (11). Indeed, a growing body of evidence suggests that people's perception of risk is mediated by social interaction. Large-scale social network analysis has shown that social interactions influence the spread of behaviors such as smoking, food choice, and adherence to various health programs (12–14). In addition, it has been found that social proximity between individuals influences both their perception of risk (15, 16) and their emotional state (17).

These findings suggest that social interaction may lead to nonlinear amplification dynamics at the scale of the population (18, 19). As illustrated by a number of studies of social contagion in relation to opinion dynamics (20), collective attention (21), cultural markets (22), and crowd behavior (23), local social influence can lead to chain reactions and amplification that yield population-scale collective patterns (24). Similarly, the dynamics of public risk perception bear the hallmarks of self-organized systems, such as opinion clustering (i.e., connected individuals share a similar judgment) and opinion polarization (i.e., opposing opinions co-exist in the same population) (25). A major challenge, therefore, is to understand what role the social transmission of risk information between individuals plays in the formation of population-level dynamics. In particular, to what extent is risk perception contagious? And what behavioral patterns might this contagion yield at the macro level?

Few theorists have attempted to analyze and model these phenomena (26). Among those that have, a common starting point is the Social Amplification of Risk (SAR) framework, which attempts to combine the cultural, structural, and psychological factors that drive the formation of risk perception at the scale of the society (27, 28). The SAR framework suggests that individuals may play the role of ''amplification stations'' by transmitting a small and often biased subset of the available information. While the overarching perspective of the SAR framework is widely assumed in risk research, it has undergone limited empirical analysis.

In this article, we study the social transmission of risk information from an interdisciplinary perspective by combining insights from the social and cognitive sciences with the study of complex systems and non-linear dynamics. In particular, we study how messages describing risk change in response to repeated social transmission by analyzing the relative fidelity with which two distinct aspects of a message — the message content and the message signal — propagate through human diffusion chains. What effect does the process of social transmission have on these message components, and what collective patterns of risk perception might this process generate?

To examine these questions we analyze how information detailing the benefits and harms of the widely used but controversial antibacterial agent *Triclosan* (29) is communicated from one individual to another in experimental diffusion chains. In a diffusion chain, a series of individuals propagate information sequentially and in turn, from one individual to the next. Specifically, the information provided by the first individual is communicated to the second individual, who in turn communicates this information to a third individual, and so on. In our experiments, participants in the chain were instructed to communicate the risks surrounding *Triclosan* in open, unstructured conversations. In each diffusion chain, the first participant was 'seeded' with information presented in media articles detailing the benefits and harms of *Triclosan*. In total, we examine 15 such diffusion chains, each composed of up to 10 participants. Both before and after participation in the experiment we also assessed the subject's risk perception of *Triclosan*. This experimental paradigm was pioneered by Frederic Bartlett over 80 years ago (30), and has since been used to study how a range of cultural entities undergo change, or cumulatively evolve, in response to repeated cultural transmission by humans and other animals (31–35). Here, we make use of the same experimental paradigm to study, for the first time, how risk information, and consequently the risk perception of people, changes when socially transmitted. We found that subjects bias the signal of the message according to their subjective perception of risk, and this influences the judgment of the receivers. Crucially, though, while risk perception "biases" propagate well and are typically amplified, the message content is transmitted with low fidelity, and tends to become shorter, gradually inaccurate, and increasingly dissimilar between chains. We use computer simulations of this process to further understand the social implications of risk amplification. Put simply, while the content of messages describing risks degenerate in response to repeated social transmission, the signal of the message propagates with high fidelity, with social transmission playing the role of an amplifying process.

## *Results*

**Message content.** We define the content of the message communicated from one participant to the next as the set of units of information that were communicated during a conversation between these participants. We identified 61 possible units of information by analyzing all conversations that occurred in the 15 diffusion chains. Each unit of information was classified in a 3-level coding scheme (see Material and Methods for the detailed procedure and examples in **Figure 1**), and tracked from one chain position to the next. This allows us to study the propagation of information down each chain. In addition, a unit of information is labeled as 'distorted' if it changes when transmitted from one chain position to the next. To illustrate, **Figure 1** depicts the propagation of risk information down a typical chain. At the first chain position, a total of 30 units of information were mentioned, 13 of which were propagated to the second chain position, and only 3 of which were propagated to the final chain position. The three successfully transmitted units were, however, distorted (as illustrated by the color-coding). In addition, 7 new units of information were generated as the chain unfolded (represented by squares), 2 of which were propagated to the end of the chain. Propagation maps for all 15 chains are shown in the Supplementary Material (**Figure S1**). As shown in **Figure 2A**, most units of information disappear as the chain unfolds, while those units that are propagated are done so with low fidelity, and tend to become increasingly distorted. In addition we measured the probability that a specific unit of information disappears from one position to the next, *pDeath,* estimated from all 15 diffusion chains. As shown in **Figure 2B**, *pDeath* has high values at the first two chain positions and reaches a relatively stable level of 0.2 afterwards. Similarly, we estimated the probability that a specific unit of information is created (*pBirth)* or gets distorted (*pDist*) from one chain position to the next. As shown in **Figure 2B**, these probabilities remain largely constant as a function of chain position. In addition, we found no significant differences when comparing these probabilities within each of the categories used in our coding scheme, which suggests that units of information (at least those that we encounter here) can appear, disappear, or undergo distortion with ostensibly constant probabilities, regardless of the kind of information communicated (**Figure S2**). Next, we analyzed how the messages develop *between* the 15 chains. Do all the chains eventually converge to a similar set of information units, or conversely, do these sets diverge from each other? To examine these questions, we measured the message differentiation coefficient $D_{ij}^p$, which defines the proportion of information units present in chain *i* at position *p* that are not present in the message of chain *j* at the same position *p*. Formally, the differentiation is defined as

$$D_{ij}^p = \frac{|m_i^p \notin m_j^p|}{|m_i^p|},$$

where $m_i^p$ is the set of information units contained in the message of chain *i* at position *p*, and $|m_i^p|$ is the size of that set. This means that $D_{ij}^p=0$ when the information units in chain *i* are also present in chain *j* at the same position (no differentiation), while $D_{ij}^p=1$ indicates that none of the information units were found in chain *j* at this position (maximum differentiation). **Figure 2C** shows the distribution of $D_{ij}^p$ for all possible pairs of messages {*i,j*} taken at position *p*=1, *p*=3 and *p*=10. The distributions tend to shift towards high differentiation values as *p* increases. While the messages communicated at the first position have an average differentiation of 0.53, this value increases to 0.87 at the 10$^{th}$ position (inset **Figure 2C**). In short, the content of the messages tends to become increasingly dissimilar between the chains as they propagate down each chain.

**Message signal.** To further understand the impact of social information transmission, we analyzed the *signal* of the message, that is, whether the message carries a positive or negative assessment of *Triclosan*. To quantify the signal of the message, we analyzed the transcripts of all conversations, and marked each sentence as '*negative*', '*positive*', or '*neutral*' (see Material and Methods). Sentences are labeled '*negative*' when they express a negative assessment of *Triclosan* (e.g., negative side effects), and labeled '*positive*' when they express a positive assessment (e.g., that *Triclosan* is safe). Sentences that express neither a positive nor a negative assessment are '*neutral*'. This coding scheme is independent of the coding scheme used to quantify the message content. We use $n_p^+$ and $n_p^-$ to denote the number of positive and negative statements found at chain position $p$, and $n_p$ the total number of sentences at position $p$. As shown in **Figure 3A**, the three quantities decrease as a function chain position, but $n_p^+$ decays faster than $n_p^-$, suggesting that negative statements propagate down the chain more freely than positive statements. Consequently, the relative proportion of negative statements $\omega_p^- = n_p^-/(n_p^+ + n_p^-)$ tends to increase gradually, at the expense of the relative proportion of positive statements $\omega_p^+ = n_p^+/(n_p^+ + n_p^-)$ (**Figure 3A** inset).

How are these aggregate trends related to the behavior of individual participants? We introduce the individual filtering coefficients $k_p^+$ and $k_p^-$ expressing the degree to which the participant at chain position $p$ modified the message (either positively or negatively) that they received from the participant at position $p$-$1$. The filtering coefficients are formally defined as $k_p^+ = n_p^+/n_{p-1}^+$, and $k_p^- = n_p^-/n_{p-1}^-$. Thus, $k_p^+ < 1$ holds for participants who attenuate the positive aspect of the message, while $k_p^+ > 1$ holds when a participant has amplified the positive aspect of the message (for negative statements we have $k_p^-$, respectively). The distribution of $k_p^+$ and $k_p^-$, across all participants, is shown in **Figure 3B**. In line with our previous finding, $k_p^-$ is on average higher than $k_p^+$ (p-value<0.001), which further underscores that our participants have an overall tendency to amplify negative statements and attenuate positive ones. However, significant individual differences were found (**Figure 3C**): While the majority of subjects tend to amplify the harmful aspects of *Triclosan* ($k_p^- > k_p^+$), other participants have the opposite profile ($k_p^+ > k_p^-$), or act neutrally on the message ($k_p^+ \approx k_p^-$). Interestingly, the individual differences can be partly explained by the risk perception $\alpha$ that subjects reported in questionnaires conducted after the experiment ($\alpha$ correlates positively with $k_p^- - k_p^+$; c=0.25, p-value=0.019). In short, the direction of the 'mutation' of the message corresponds to the opinion of the speaker: individuals with higher risk perception tend to filter out positive statements and amplify the dangerous aspects of *Triclosan*, while those with lower risk perception tend to have an opposite effect on the signal of the message.

Finally, we measured what impact a message has on the risk perception of the receiver. Overall, participants changed their risk perception by an absolute average value of $|\Delta_\alpha| = |\alpha - \alpha^0| = 0.19$ (STD=0.17), where $\alpha^0$ is the risk perception of the individual before the experiment, and $\alpha$ is the risk perception of the same subject after the experiment. The degree of change $\Delta_\alpha$ should be interpreted relative to the signal $\sigma_p$ of the message that this participant received, defined as $\sigma_p = \omega_p^-$: When the message signal confirms the prior assessment of the participant (i.e. $\sigma_p \approx \alpha_0$), no change of opinion is expected (i.e. $\Delta_\alpha \approx 0$), whereas $\Delta_\alpha$ should vary in the same direction as $\sigma_p - \alpha_0$. As shown in **Figure 4**, $\Delta_\alpha$ and $\sigma_p - \alpha_0$ correlate well (correlation c=0.55, p-value<0.001), confirming that participants are influenced by the signal of the message they receive.

**Collective patterns.** These findings provide quantitative insights into the role of social transmission as an amplifying process. Specifically, the signal of the incoming message influences the receiver's risk perception, which in turn shapes the signal of the outgoing message. In many social systems, similar feedback loops are typically associated with non-linear dynamics and emerging collective patterns (36). In our experiment, the social amplification of the risk signal is clearly visible (**Figure 3A**). However, considerable behavioral fluctuations are observed within and between chains, making it difficult to conduct a systematically controlled analysis of the amplification process. To further examine this amplifying process, we analyzed a simple simulation model derived from our observations to study this amplification process more precisely.

In the model, each individual $i$ in a chain is has an initial risk perception $\alpha_i^0$. Individuals sequentially, and in turn, receive and transmit a message made of $n^+$ and $n^-$ positive and negative statements, respectively. The signal $\sigma$ of the message is defined as above, by the equation $\sigma = n^-/(n^+ + n^-)$. When an individual $i$ receives the message, the risk level of that individual changes as follows:

$$\alpha_i = \alpha_i^0 + s\,(\sigma - \alpha_i^0), \qquad [1]$$

where *s* is an influence factor with a value between between 0 (no influence) and 1 (full adoption). Previous empirical estimates of influence factors lie between 0.3 and 0.5 (37, 38), which is consistent with our experimental estimate ($s=0.45$, **Figure 4**). In our simulation, when a message is transmitted from one person to another, it undergoes a random, biased mutation dependent on the risk perception $\alpha_i$ of the emitter $i$. For the sake of simplicity, we implement the message mutation such that $n^+$ is increased by 1 with a probability $p_i^+ = \alpha_i$, and decreased by 1 with a probability $p_i^- = 1 - \alpha_i$. Similarly, $n^-$ is increased by 1 with the probability $p_i^-$, and decreased by 1 with the probability $p_i^+$. Based on this simple model, we implemented a series of computer simulations.

While the impact of the message on the first person of the chain is given by equation [1], the dynamics at subsequent chain positions is complex and results from repeated mutations of the message combined with related changes in the judgment of the individuals. First, we conducted simulations under conditions matching those of our experiment, where there exists a substantial diversity of initial risk perceptions (i.e. where $\alpha_i^0$ is drawn from a normal distribution with mean 0.65 and standard deviation 0.22). At chain position $p=10$, the aggregate distributions of risk signals and risk perceptions correspond to those we observed experimentally (ks-test p=0.57 and p=0.08). As we observed previously, the model predicts a gradual amplification of the risk signal, whereas individuals' final risk perception exhibits a considerable variability due to the diversity of initial judgments (see simulation examples in **Figure S3**). Under homogenous initial conditions, where all individuals have the same initial risk perception (i.e. $\alpha_i^0 = \alpha^0$ for all individuals $i$), the amplification process is itself amplified and easier to study. For example, our simulations show that a neutral message (i.e. $n^+=n^-$) injected in a population of concerned individuals (i.e. $\alpha^0 = 1$ for all individuals in the chain) tends to mutate rapidly as it propagates from one person to another, culminating in a steady-state representing the population's initial view, with no impact on the individuals opinions in the long run (**Figure S3B**). What social dynamics emerge when the initial risk perception $\alpha^0$ of the group, the signal $\sigma$ of the injected message, and the social influence factor $s$ are varied? **Figure 5** and **Figure S4** examine such an exploration of the model's parameter space. In short, the social amplification of the risk signal is visible after 10 transmissions, and continues to increase as the message propagates further (**Figure 5A**). After 50 transmissions, the signal of the message is either strongly positive (dark blue) or strongly negative (dark red). The impact on individuals' risk perception follows a similar trend (**Figure 5B**). In most cases, the message has a polarizing effect on the population, where people's biased initial judgments tend to become more extreme at the end of the chain, even when the injected message is completely neutral. Along the transition zone depicted in green, the mean values of the signal and the risk perception are both 0.5, but these means have very high standard deviations, indicating a U-shaped distribution where the outcome can be either high or low (**Figure S5**). Thus, even a neutral message injected into a neutral population can possibly have a polarizing effect on people's judgment due to the amplification of random mutations of the message (as illustrated **Figure S3C**).

## *Discussion*

Among the factors influencing public risk perception the impact of social transmission is arguably the least studied, despite being an everyday occurrence in an increasingly connected society. This study is the first to examine experimentally how the social transmission of risk information though human subjects influences the propagation of risky information, and how these risk perceptions become amplified as a result of this propagation. Ambiguity surrounds the risks associated with *Triclosan,* such that each message not only carries factual information about the potential benefits and harms, but also a signal representing the subjective judgment of the communicator. Consequently, we have distinguished between *content* propagation and *signal* propagation. First, we found that the information content of a message degrades (contains fewer units information) and becomes less accurate (undergoes distortion) in response to being transmitted. As a consequence, different communication chains led to a focus on different issues related to *Triclosan*, including Greenpeace protests (chain 13), breast-feeding (chain 10), and environmental damage (chain 3). Participants nevertheless influence each other: changes in risk perception are a function of the signal of the received message (whether the message is positive of negative). Thus, changes in risk perception occur less as a result of the message content, and more as a result of its overarching, subjective signal. In agreement with Bartlett's experiments (30), we observed that the information passed along the chain is shaped by the preconceptions of its members. In Bartlett's experiment, a Native American story was gradually modified to fit the cultural perceptions of the British participants. In our experiment, a similar mechanism drives the propagation of, and modifications to, "stories" detailing the risks associated with *Triclosan*: Participants tended to alter the signal of the message to make it closer to their initial judgment.

At the aggregate level, the process of social transmission tends to amplify or attenuate the message signal. As a message propagates down the chain, it becomes distorted to fit the view of those transmitting it, with the original message eventually having a negligible impact on the judgment of the participants. Ultimately, the message can have a counter-intuitive polarizing effect on the population: After a few rounds of distortion, the message can strengthen the existing bias of the group, even though it initially supported the opposite view. As our simulations show, this social phenomenon is itself amplified in chains of like-minded individuals. According to recent theoretical and empirical studies (25, 39, 40), social interactions and information exchange frequently take place within clusters of like-minded people, suggesting that "natural" social structures may create favorable conditions for strong amplification dynamics.

More generally, our experiments demonstrate how the field of opinion dynamics can be extended to include the study of risk perception, and how existing theoretical models of collective risk perception can be furnished with empirical support (39, 41, 42). Perhaps most importantly, our findings illustrate the importance of studying patterns of risk perception as, at least in part, the outcome of a social process. The role of complex networks (43) with highly connected individuals (44) on the development of behavioral patterns of risk perception, however, remains an open question, as does the role of private, public, and social information (45) in driving, or curtailing, the propagation of information describing risks.

## Material and Methods

**Experimental design**. The experiment took place in April 2013. We invited 12 participants to each of the 15 experimental sessions. Due to absences, some sessions were conducted with less than 12 participants. In total, we collected data from 4 chains of length 12, 2 chains of length 11, 4 chains of length 10, 3 chains of length 9, 1 chain of length 8, and 1 chain of length 7 (average chain size 10.1). Our analyses considered only the first 10 participants of diffusion chains of length greater than 10. All participants gave informed consent to the experimental procedure, and received a flat fee of 15 euro. The participants entered a waiting room and were instructed not to interact with each other. All participants completed an initial questionnaire Q1. The first participant $p_1$ was then moved to the experimental room and instructed to read a collection of 6 media articles displayed on a computer screen. The articles presented alternative views on the use and the suspected side effects of the controversial antibacterial agent *Triclosan* (these articles are provided in the Supplementary Material). In order to present a representative selection of articles, and one that reflects those likely to be encountered in an everyday setting, we selected each article from the first page of results returned by Google with the search keyword '*Triclosan*' (accessed early 2013). The articles were presented in a random order across groups, and the subject $p_1$ had 3 minutes to read each of them.

After the reading phase (18 minutes in total), the computer was shut down, and participant $p_2$ was invited to join participant $p_1$ in the experimental room. Both participants were instructed to discuss *Triclosan* in an open, unstructured discussion. No time limit was imposed. At the end of the discussion, the next participant $p_3$ was moved to the experimental room and a new discussion began between $p_2$ and $p_3$ under the same conditions. In the meantime, the first participant $p_1$ was instructed to fill the second questionnaire Q2 in another room, and was then free to leave. This procedure was repeated for all participants in the chain. The discussions were recorded, and the audio files were then transcribed and translated into English (from German).

**Measures of risk perception.** The risk perception of the participants towards *Triclosan* was measured before and after the discussion by means of two questionnaires Q1 and Q2. Questionnaire Q1 evaluated the subject's knowledge about *Triclosan*. Among all participants, only one had prior knowledge of *Triclosan*, but failed to answer basic questions relating to *Triclosan*. We therefore considered all subjects to be unfamiliar with *Triclosan*. Questionnaires Q1 and Q2 also evaluated the risk perception of the participants towards *Triclosan*. The risk perception of the participants was self-reported by placing tick on a continuous line ranging from '*Not dangerous at all*', to '*Extremely dangerous*'. Because participants were unfamiliar with *Triclosan* before the experiment, it was impossible to assess their level of risk perception directly. We therefore used an indirect measurement by evaluating their perception of risk toward the more general issue of chemical use in food safety and cosmetics.

**Coding the content of the message.** We first used the transcripts of the discussions to identify every unit of information that was discussed. We identified a list of 61 units of information. These units were classified into 4 high-level categories: '*Side effects*', '*Personal anecdotes*', '*Where is Triclosan*', and '*Others*'. Each of these high-level categories was then subdivided into sub-categories and sub-sub-categories. For example, the top category '*Side effects*' contains the sub-category '*In mice*', which further divides into 7 classes such as '*Heart diseases*' and '*Cancer*'. Finally, we analyzed the transcripts and tracked each unit of information from one chain position to the next, until the end of each chain. In addition, each unit of information was labeled '*distorted*' if differences or imprecisions were detected from one chain position to the next.

**Coding the signal of the message.** We also used the transcripts to analyze the signal of the conversation. Here, we reviewed all sentences pronounced by the informed speaker at each chain position, and labeled each of them as '*Positive*, '*Negative*, or '*Neutral*'. Sentences that highlight the suspected dangers of *Triclosan* or express a negative judgement about it received a '*Negative*' label, while those suggesting that the use of *Triclosan* is safe or well-controlled received a '*Positive*' label. Sentences that were labeled neither '*Negative* nor ''*Positive*' were considered '*Neutral*'. This procedure was conducted by two independent coders, who received the same sheet of instructions (available in the Supplementary Material). The results were then averaged.

## *Acknowledgements*

We thank Eleonora Spanudakis and Mareike Trauernicht for helpful assistance during data collection and data analyses. We are grateful to Christel Fraser for her help in transcribing and translating the audio recordings. We thank Michael Mäs, Jan Lorenz, and Kenny Smith for insightful discussions. The authors would also like to acknowledge the members of the Center for Adaptive Rationality and the Center for Adaptive Behavior and Cognition at the Max Planck Institute for Human Development for providing valuable feedback during the preparation of this work.

## *References*

1. Slovic P (2000) *The Perception of Risk* (Earthscan Publications, London).

2. Slovic P (1987) Perception of risk. *Science* 236(4799):280–285.

3. Funk S, Gilad E, Watkins C, Jansen V (2009) The spread of awareness and its impact on epidemic outbreaks. *Proc Natl Acad Sci USA* 106(16):6872–6877.

4. Huang L, et al. (2013) Effect of the Fukushima nuclear accident on the risk perception of residents near a nuclear power plant in China. *Proc Natl Acad Sci USA* 110(49):19742–19747.

5. Loewenstein G, Weber E, Hsee C, Welch N (2001) Risk as Feelings. *Psychol Bull* 127(2):267–286.

6. Pachur T, Hertwig R, Steinmann F (2012) How do people judge risks: availability heuristic, affect heuristic, or both? *J Exp Psychol Appl* 18(3):314–330.

7. Slovic P, Peters E (2006) Risk Perception and Affect. *Curr Dir Psychol Sci* 15(6):322–325.

8. Fischhoff B, Brewer NT, Downs JT (2011) *Communicating Risks and Benefits: An Evidence-Based User's Guide* (Silver Spring, MD: U.S. Food and Drug Administration).

9. Gigerenzer G, Gaissmaier W, Kurz-Milcke E, Schwartz L, Woloshin S (2007) Helping Doctors and Patients Make Sense of Health Statistics. *Psychol Sci Public Interes* 8(2):53–96.

10. Faria J, Krause S, Krause J (2010) Collective behavior in road crossing pedestrians: the role of social information. *Behav Ecol* 21(6):1236–1242.

11. Salathé M, Vu D, Khandelwal S, Hunter D (2013) The dynamics of health behavior sentiments on a large online social network. *EPJ Data Sci* 2(1):1–12.

12. Christakis N, Fowler J (2007) The Spread of Obesity in a Large Social Network over 32 Years. *N Engl J Med* 357(4):370–379.

13. Christakis N, Fowler J (2008) The Collective Dynamics of Smoking in a Large Social Network. *N Engl J Med* 358(21):2249–2258.

14. Centola D (2010) The Spread of Behavior in an Online Social Network Experiment. *Science* 329(5996):1194–1197.

15. Scherer C, Cho H (2003) A Social Network Contagion Theory of Risk Perception. *Risk Anal* 23(2):261–267.

16. Binder A, Scheufele D, Brossard D, Gunther A (2011) Interpersonal Amplification of Risk? Citizen Discussions and Their Impact on Perceptions of Risks and Benefits of a Biological Research Facility. *Risk Anal* 31(2):324–334.

17. Kramer A, Guillory J, Hancock J (2014) Experimental evidence of massive-scale emotional contagion through social networks. *Proc Natl Acad Sci USA* 111(24):8788–8790.

18. Schelling T (1978) *Micromotives and Macrobehavior* (W. W. Norton).

19. Milgram S (1977) *The Individual in a Social World: Essays and Experiments* (Addison Wesley Publishing Company).

20. Lorenz J, Rauhut H, Schweitzer F, Helbing D (2011) How social influence can undermine the wisdom of crowd effect. *Proc Natl Acad Sci USA* 108(22):9020–9025.

21. Wu F, Huberman B (2007) Novelty and collective attention. *Proc Natl Acad Sci USA* 104(45):17599–17601.

22. Salganik M, Dodds P, Watts D (2006) Experimental study of inequality and unpredictability in an artificial cultural market. *Science* 311(5762):854–856.

23. Moussaïd M, Helbing D, Theraulaz G (2011) How simple rules determine pedestrian behavior and crowd disasters. *Proc Natl Acad Sci USA* 108(17):6884–6888.

24. Helbing D, et al. (2014) Saving Human Lives: What Complexity Science and Information Systems can Contribute. *J Stat Phys* DOI 10.100.

25. Salathé M, Khandelwal S (2011) Assessing Vaccination Sentiments with Online Social Media: Implications for Infectious Disease Dynamics and Control. *PLoS Comput Biol* 7(10):e1002199.

26. Moussaïd M (2013) Opinion Formation and the Collective Dynamics of Risk Perception. *PLoS One* 8(12):e84592.

27. Renn O, Burns W, Kasperson J, Kasperson R, Slovic P (1992) The Social Amplification of Risk: Theoretical Foundations and Empirical Applications. *J Soc Issues* 48(4):137–160.

28. Kasperson R, et al. (1988) The social amplification of risk: a conceptual framework. *Risk Anal* 8(2):177–187.

29. Yueh M-F, et al. (2014) The commonly used antimicrobial additive triclosan is a liver tumor promoter. *Proc Natl Acad Sci USA* 111(48):17200–17205.

30. Bartlett F (1932) *Remembering: A Study in Experimental and Social Psychology* (Cambridge University Press).

31. Kempe M, Mesoudi A (2014) Experimental and theoretical models of human cultural evolution. *WIREs Cogn Sci* 5(3):317–326.

32. Griffiths T, Kalish M, Lewandowsky S (2008) Theoretical and empirical evidence for the impact of inductive biases on cultural evolution. *Philos Trans R Soc Lond B Biol Sci* 363(1509):3503–3514.

33. Horner V, Whiten A, Flynn E, de Waal F (2006) Faithful replication of foraging techniques along cultural transmission chains by chimpanzees and children. *Proc Natl Acad Sci USA* 103(37):13878–13883.

34. Kirby S, Cornish H, Smith K (2008) Cumulative cultural evolution in the laboratory: An experimental approach to the origins of structure in human language. *Proc Natl Acad Sci USA* 105(31):10681–10686.

35. Mesoudi A, Whiten A, Laland K (2006) Towards a unified science of cultural evolution. *Behav Brain Sci* 29(4):329–347.

36. Helbing D (1995) *Quantitative Sociodynamics: Stochastic Methods and Models of Social Interaction Processes* (Kluwer Academic).

37. Soll J, Larrick R (2009) Strategies for revising judgment: how (and how well) people use others' opinions. *J Exp Psychol Learn Mem Cogn* 35(3):780–805.

38. Moussaid M, Kaemmer J, Analytis P, Neth H (2013) Social Influence and the Collective Dynamics of Opinion Formation. *PLoS One* 8(11):e78433.

39. Mäs M, Flache A, Helbing D (2010) Individualization as driving force of clustering phenomena in humans. *PLoS Comput Biol* 6(10):e1000959.

40. Lorenz J (2007) Continuous opinion dynamics under bounded confidence: A survey. *Int J Mod Phys* 18(12):1819–1838.

41. Castellano C, Fortunato S, Loreto V (2009) Statistical physics of social dynamics. *Rev Mod Phys* 81(2):591–646.

42. Deffuant G, Neau D, Amblard F, Weisbuch G (2001) Mixing beliefs among interacting agents. *Adv Complex Syst* 3:87–98.

43. Watts D, Strogatz S (1998) Collective dynamics of "small-world" networks. *Nature* 393(6684):440–442.

44. Watts D, Dodds P (2007) Influentials, Networks, and Public Opinion Formation. *J Consum Res* 34(4):441–458.

45. Gallup A, et al. (2012) Visual attention and the acquisition of information in human crowds. *Proc Natl Acad Sci USA* 109(19):7245–7250.
17

*Figures*

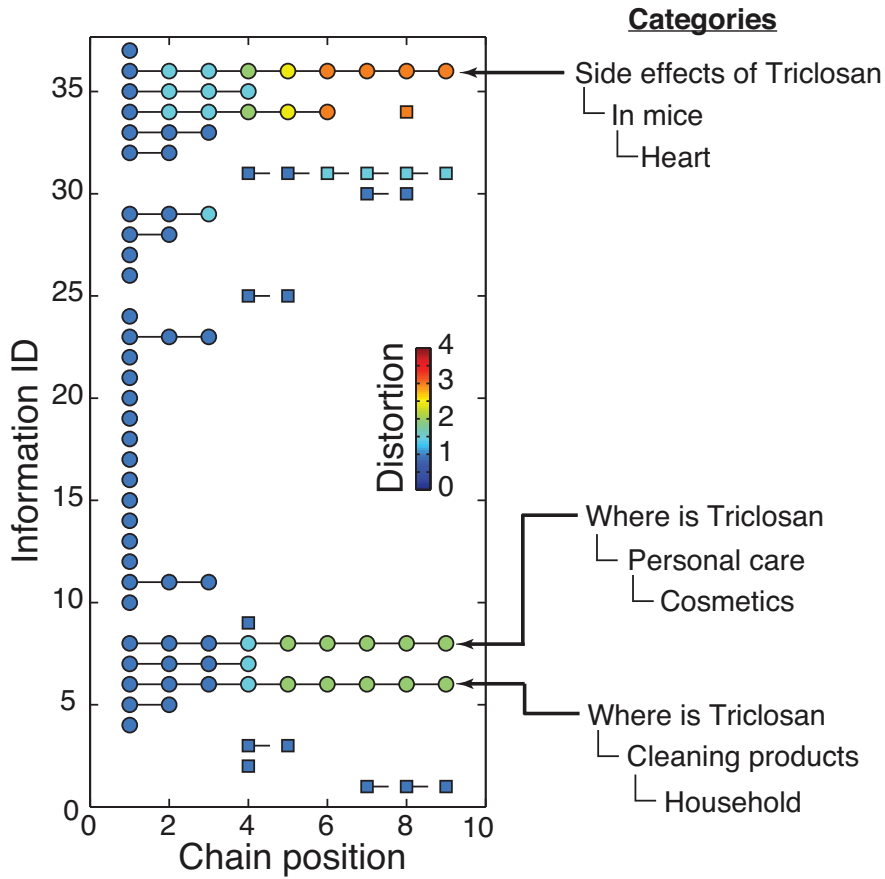

**Figure 1**: A topological map of information propagation in an experimental diffusion chain. Among all units of information available at chain position 1 (blue dots), only 3 have survived to the end of the chain, although they were strongly distorted. The text on the right-hand side describes the categories of these units of information. Seven units of information were introduced by the chain (squares), two of which were survived to the end of the chain. The color-coding indicates the cumulated distortion of the information. Information IDs (y-axis) are arbitrary.

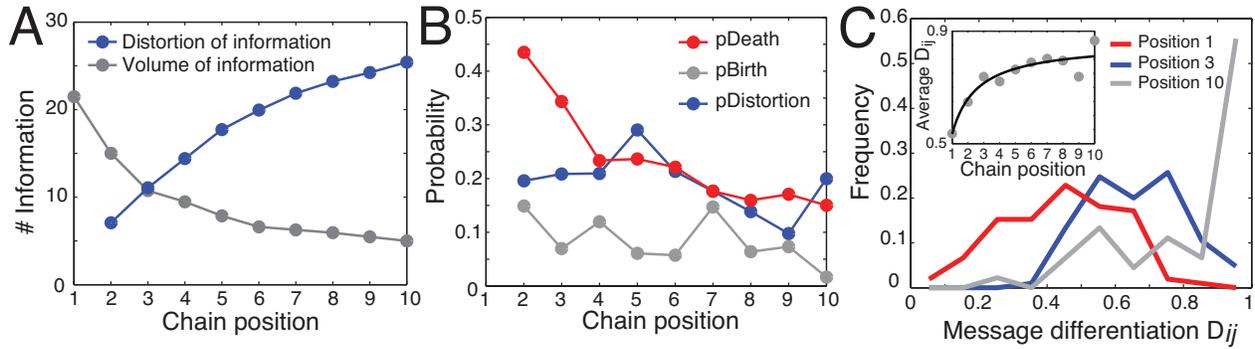

**Figure 2**: The dynamics of information propagation. (A) The mean number of units of information transmitted and the mean cumulated distortion of the information over the chains. (B) The hazard functions showing the probabilities that a piece of information disappears (red), gets distorted (blue), or appears (grey) at each chain position. (C) The distribution of the message differentiation $D_{ij}$ for all possible pairs of messages $i$ and $j$ at position 1 (red line), position 3 (blue line), and position 10 (grey line). The distributions tend to shift towards high differentiation values, indicating that the content of the message becomes increasingly different between the chains as it propagates from one person to another. The inset indicates the mean value of $D_{ij}$ at each position of the chain (grey dots). The dark line corresponds to the fit equation $f(x) = p_1 x/(x + p_2)$ with parameters $p_1$=0.86 and $p_2$=0.61.

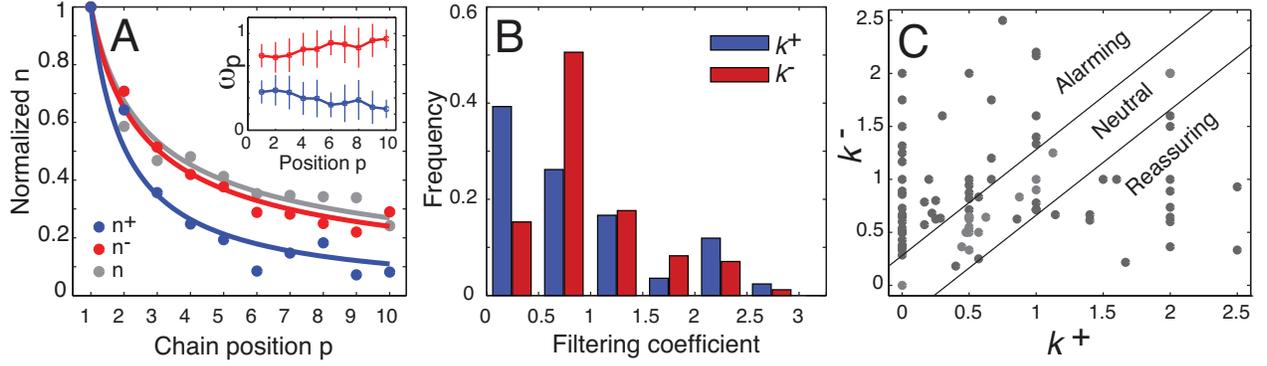

**Figure 3**: Message mutation. (A) The evolution of the normalized number of positive statements $n_p^+$ (blue), negative statements $n_p^-$ (red), and total statements $n_p$ (grey) over the chain. Fit lines are power functions $f(x) = x^{-e}$, with the exponent $e$ equals 0.96, 0.62, and 0.57 for $n_p^+$, $n_p^-$, and $n_p$, respectively. The inset indicates the average proportion of positive statements ($\omega_p^+$, in blue) and negative statements ($\omega_p^-$, in red) along the chains. (B) The distribution of the filtering coefficients $k_p^+$ and $k_p^-$ for all subjects. The distribution of $k_p^-$ is significantly higher than $k_p^+$ (p-value<0.001). (C) Individual profiles measured as the pair $\{k_p^+, k_p^-\}$. Each point represents one experimental subject. Individuals with $k_p^+ \approx k_p^-$ have a neutral effect on the message, whereas those with $k_p^- > k_p^+$ (resp. $k_p^+ > k_p^-$) tend to make the message more alarming (resp. more reassuring).

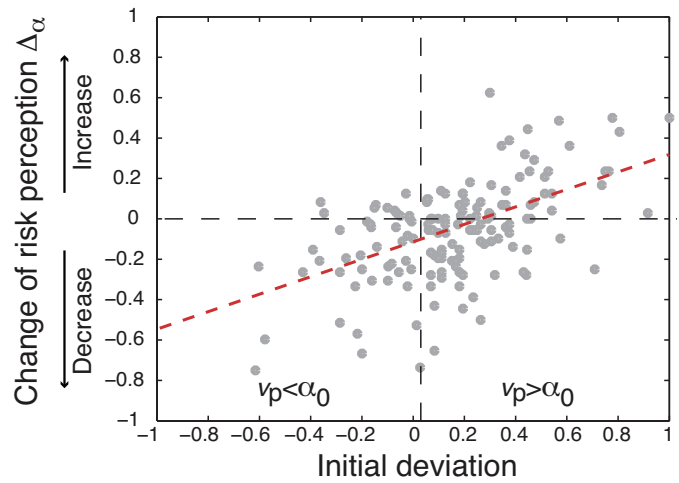

**Figure 4**: Social influence. Scatter plot representing the degree to which participants changed their risk perception as a function of the initial deviation with the message signal they have received. Subjects receiving a message with negative signal, with respect to their initial perception, tend to increase their risk level, while those receiving a more positive signal tend to reduce their risk level. The equation of the fit line is $y = 0.45x - 0.14$.

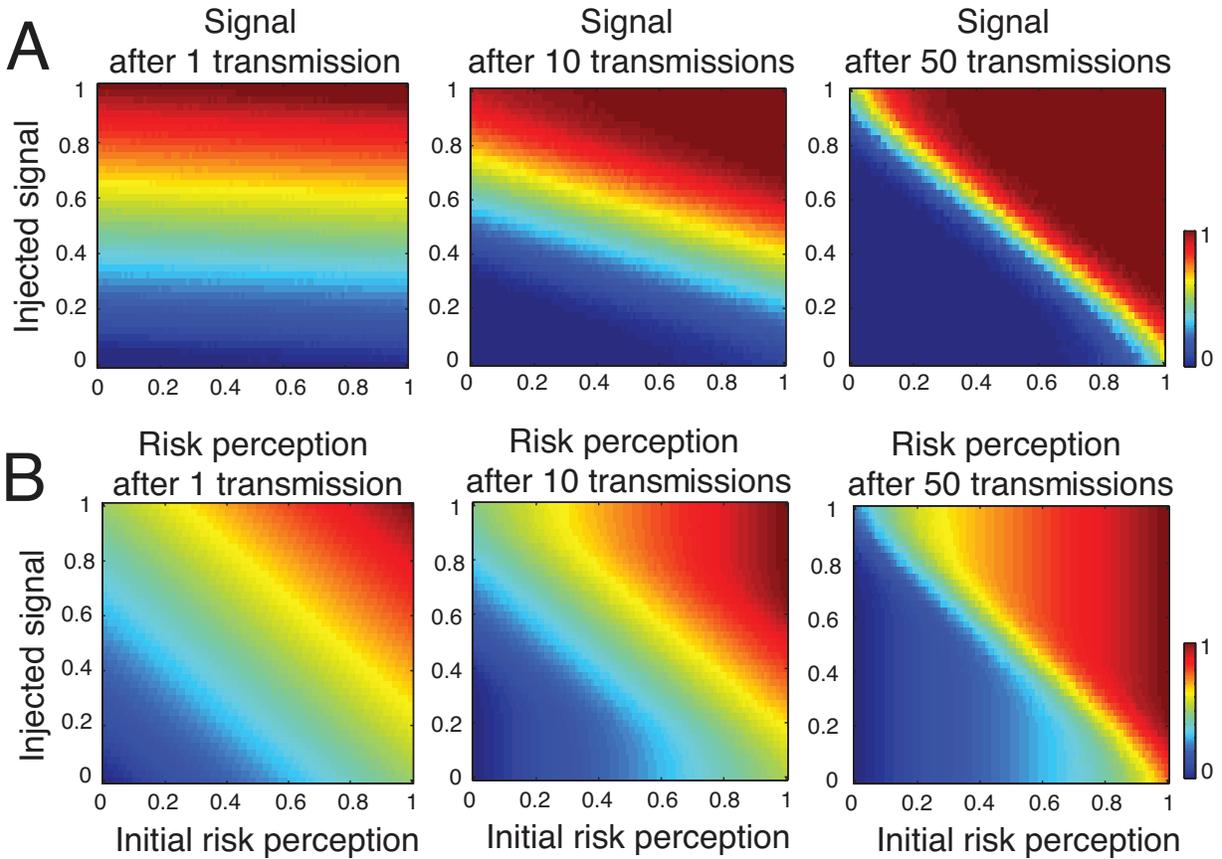

**Figure 5**: Computer simulations. (A) The evolution of the risk signal at chain positions 1, 10 and 50, as a function of the initial risk perception of the individuals (x-axes) and the signal of the injected message (y-axes). The gradual dominance of extreme values (in dark red and dark blue) demonstrates the amplification of the risk signal. (B) The evolution of individuals' risk perception follows a similar trend. Individuals who initially expressed extreme risk perception (x=0 or x=1) gradually move back to their initial view, regardless of the message signal. The social influence parameter is set to *s*=0.5. Simulations varying the value of *s* are provided in the Supplementary Information (**figure S4**). Results are averaged over 500 simulations.

# The amplification of risk in experimental diffusion chains

Supporting Information

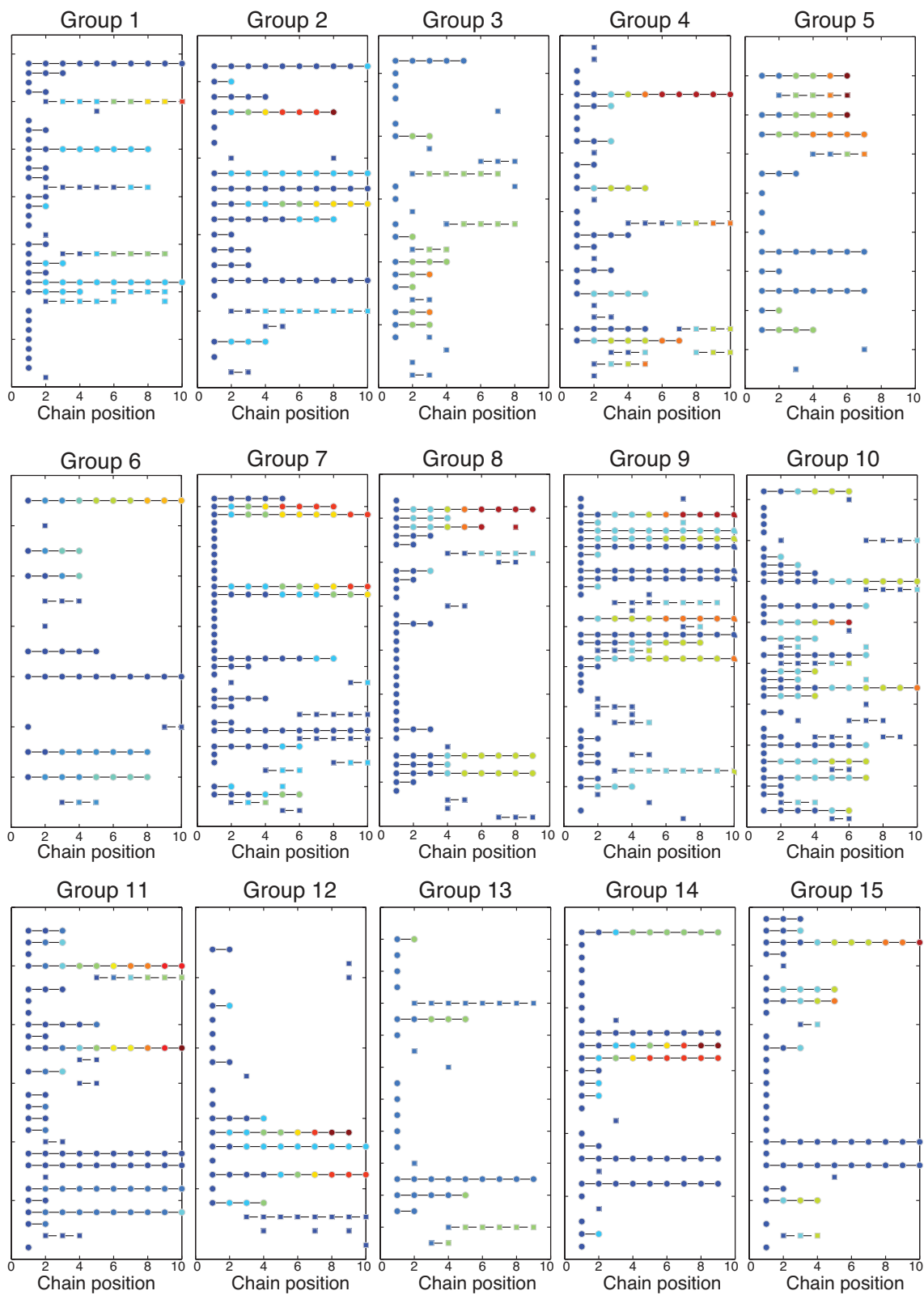

**Figure S1**: Topological maps of information propagation for all 15 replications of the experiment. Group 8 is shown and commented in the main text (**Figure 1**).

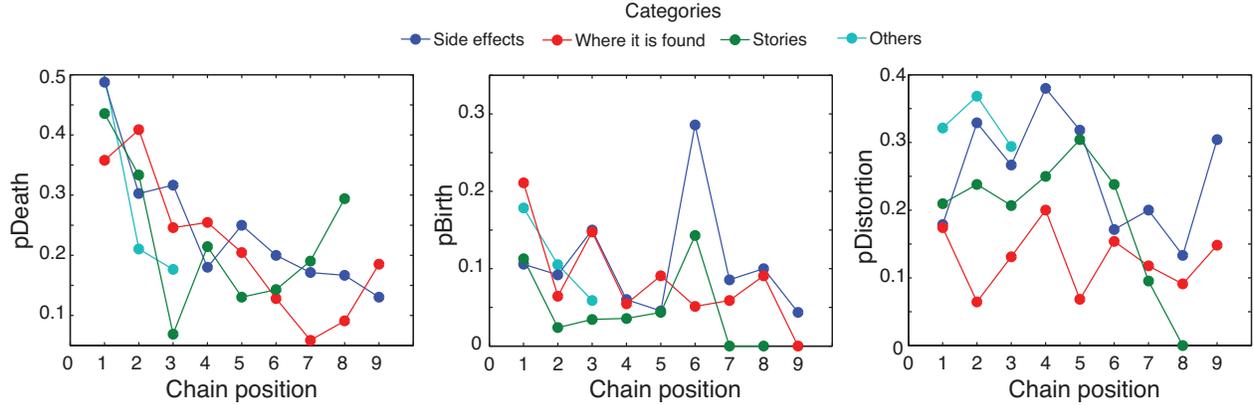

**Figure S2**: Hazard functions showing the probabilities that a piece of information disappears (left), appears (middle), or gets distorted (right) at each chain position, compared between types of information. The four lines correspond to four different categories of information indicated on the top of the figure. No specific difference is found between categories, suggesting that any pieces of information can appear, disappear, or get distorted, regardless of the nature of the information that is communicated.

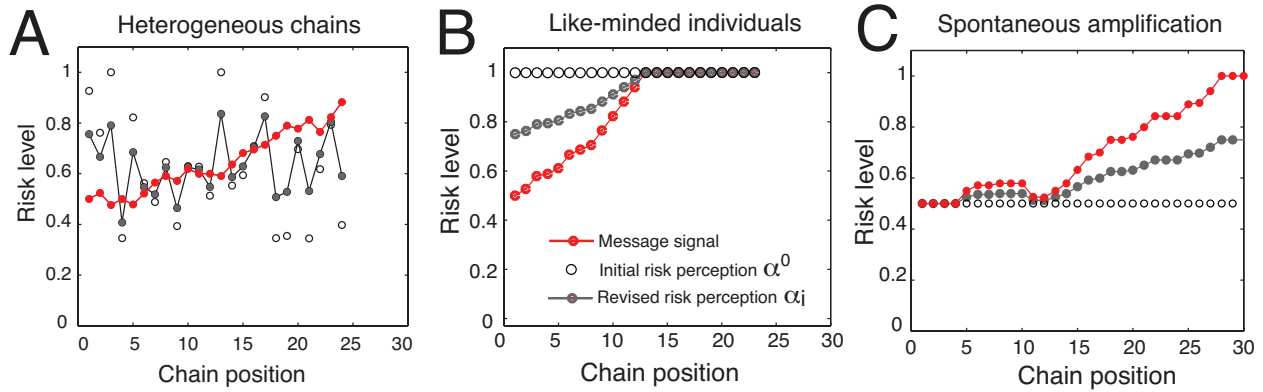

**Figure S3**: Representative examples of model simulations. In (A) people composing the chain express different initial opinions. Heterogeneity of initial judgments is based on the values observed in our experiment (average value 0.65 and standard deviation 0.22). As observed in our data, the model predicts a gradual amplification of the risk signal (red curve) and a considerable variability of revised judgments, which reflect the initial diversity risk perceptions. (B) In chains of like-minded people, the amplification process is exacerbated. As the message is communicated from one person to another, it gradually mutates to fit the initial view of the population. At the end of the chain, individuals are no longer influenced by the injected message. (C) Illustration of a spontaneous amplification case, where a neutral message injected in a neutral chain generates high levels of risk perceptions. This phenomenon results from the amplification of random mutations of the signal. In all 3 figures, the injected message is neutral with $n^+ = 10$ positive statements, and $n^- = 10$ negative statements (and thus signal $\sigma=0.5$).

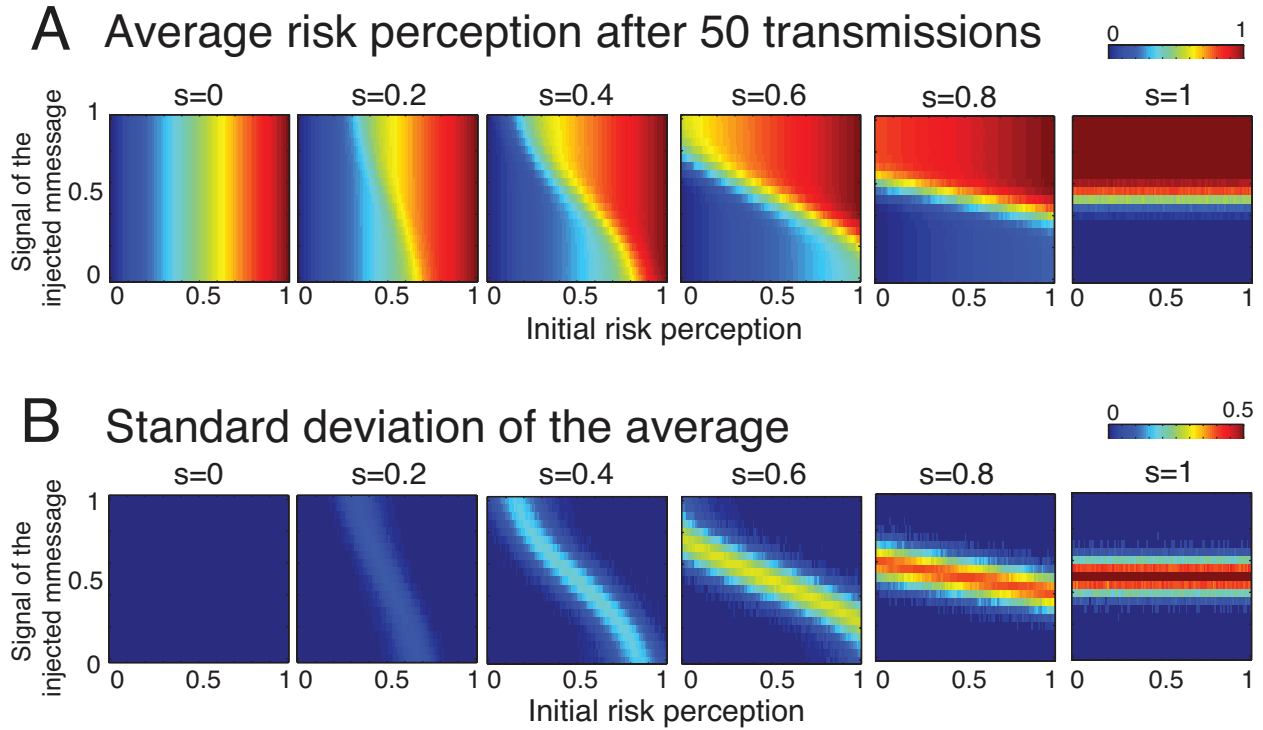

**Figure S4**: Effects of varying the social influence coefficient *s* on the model's predictions. At *s*=0, social influence is inexistent: The final opinion of the individuals equals its initial value (A-left) and the system is deterministic (B-left). As *s* increases, the S-shaped transition zone emerges and becomes gradually sharper (A). At the transition zone, the system becomes increasingly unpredictable (B). At *s*=1, social influence is at the maximum level, and individuals fully adhere to the risk message, regardless of their initial opinion. This latter case exhibits strong amplification effects, where any message with signal lower than 0.5 converges to 0, whereas a message with signal higher than 0.5 converges to 1 (A-right). Neutral messages carrying a signal around 0.5 can equally tip up to 1 or down to 0 (B-right).